\documentclass{emulateapj}

\def\msun{M$_{\odot}$}

\def\asec{\ifmmode ^{\prime\prime}\else$^{\prime\prime}$\fi}
\def\degs{\ifmmode ^{\circ}\else$^{\circ}$\fi}
\def\amin{\ifmmode ^{\prime}\else$^{\prime}$\fi}
\def\gtrsim{\mathrel{\hbox{\rlap{\hbox{\lower4pt\hbox{$\sim$}}}\hbox{$>$}}}}
\def\lessim{\mathrel{\hbox{\rlap{\hbox{\lower4pt\hbox{$\sim$}}}\hbox{$<$}}}}
\def\msun{M$_{\odot}$}

\shorttitle{PTF Science}
\shortauthors{A.~Rau et al.}

\begin{document}


\title{Exploring the Optical Transient Sky with the Palomar Transient Factory}


\author{Arne Rau\altaffilmark{1,2}, Shrinivas~R. Kulkarni\altaffilmark{1}, Nicholas~M. Law\altaffilmark{1}, Joshua~S. Bloom\altaffilmark{3}, David Ciardi\altaffilmark{4}, George~S. Djorgovski\altaffilmark{1}, Derek~B. Fox\altaffilmark{5}, Avishay Gal-Yam\altaffilmark{6}, Carl~C. Grillmair\altaffilmark{7},  Mansi~M. Kasliwal\altaffilmark{1}, Peter~E. Nugent\altaffilmark{8}, Eran~O. Ofek\altaffilmark{1}, Robert~M. Quimby\altaffilmark{1}, William~T. Reach\altaffilmark{9}, Michael~Shara\altaffilmark{10},  Lars Bildsten\altaffilmark{11},  S.~Bradley Cenko\altaffilmark{3}, Andrew~J. Drake\altaffilmark{1}, Alexei~V. Filippenko\altaffilmark{3},  David~J. Helfand\altaffilmark{12}, George Helou\altaffilmark{9}, D.~Andrew Howell\altaffilmark{13,14},  Dovi Poznanski\altaffilmark{3,8}, Mark Sullivan\altaffilmark{15}}

\affil{$^1$ Caltech Optical Observatories, MS 105-24, California Institute of Technology, Pasadena, CA 91125, USA}
\affil{$^2$ Max-Planck Institute for extraterrestrial Physics, Garching, 85748, Germany}
\affil{$^3$ Department of Astronomy, University of California, Berkeley, CA 94720-3411, USA}
\affil{$^4$ Michelson Science Center, MS 100-22, California Institute of Technology, Pasadena, CA 91125, USA}
\affil{$^5$ Department of Astronomy and Astrophysics, Pennsylvania State University, University Park, PA 16802, USA}
\affil{$^6$ Benoziyo Center for Astrophysics, Weizmann Institute of Science, 76100 Rehovot, Israel}
\affil{$^7$ Spitzer Science Center, MS 220-6, California Institute of Technology,  Pasadena, CA 91125, USA}
\affil{$^8$ Lawrence Berkeley National Laboratory, Berkeley, CA 94720, USA}
\affil{$^{9}$ Infrared Processing and Analysis Center, California Institute of Technology, MS 100-22, Pasadena, CA 91125, USA}
\affil{$^{10}$ Department of Astrophysics, American Museum of Natural History, New York, NY 10024, USA}
\affil{$^{11}$ Kavli Institute for Theoretical Physics and Department of Physics, University of California, Santa Barbara, CA 93106, USA}
\affil{$^{12}$ Columbia Astrophysics Laboratory, Columbia University, New York, NY 10027, USA}
\affil{$^{13}$ Las Cumbres Global Telescope Network, 6740 Cortona Dr. Santa Barbara, CA 93117, USA}
\affil{$^{14}$ University of California, Santa Barbara, CA 93106, USA}
\affil{$^{15}$ Department of Physics (Astrophysics), University of Oxford, Denys Wilkinson Building, Keble Road, Oxford, OX1 3RH, UK}

\email{arne@astro.caltech.edu}

\begin{abstract}
  The  Palomar  Transient Factory  (PTF)  is  a wide-field  experiment
  designed to  investigate the optical  transient and variable  sky on
  time  scales from  minutes  to  years. PTF  uses  the CFH12k  mosaic
  camera, with  a field of view  of 7.9\,deg$^2$ and a  plate scale of
  1\asec\ pixel$^{-1}$, mounted on the the Palomar Observatory 48-inch
  Samuel Oschin  Telescope.  The PTF operation strategy  is devised to
  probe the existing  gaps in the transient phase  space and to search
  for theoretically  predicted, but not yet  detected, phenomena, such
  as fallback  supernovae, macronovae,  .Ia supernovae and  the orphan
  afterglows  of gamma-ray bursts.   PTF will  also discover  many new
  members of known source classes, from cataclysmic variables in their
  various avatars  to supernovae and active galactic  nuclei, and will
  provide  important  insights  into understanding  galactic  dynamics
  (through  RR~Lyrae  stars)  and  the  Solar  system  (asteroids  and
  near-Earth objects). The  lessons that can be learned  from PTF will
  be  essential  for the  preparation  of  future  large synoptic  sky
  surveys like the Large Synoptic  Survey Telescope.  In this paper we
  present the scientific motivation for PTF and describe in detail the
  goals and expectations for this experiment.

\end{abstract}
\keywords{Supernovae, Quasars and Active Galactic Nuclei, Stars, Extrasolar Planets}

\section{Introduction}
\label{sec:introduction}

\begin{figure}
\begin{center}
\centering
\includegraphics[width=0.40\textwidth,angle=90]{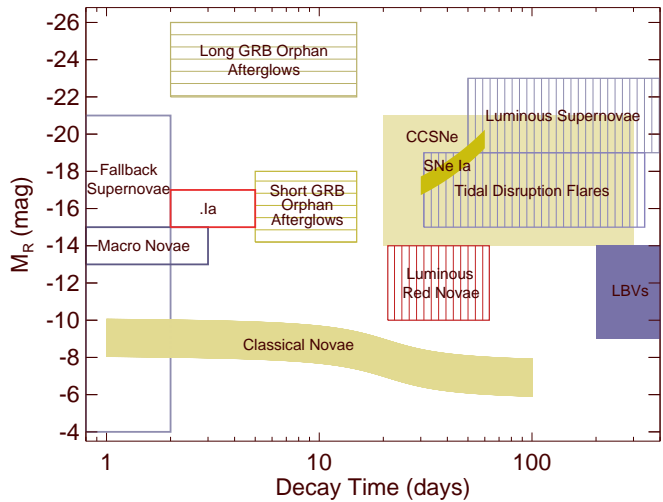}
\end{center}
\caption{$R$-band peak  magnitude as a function  of characteristic decay
  timescale (typically the  time to fade from peak  by 2 magnitudes)
  for luminous  optical transients  and variables.  Filled  boxes mark
  well-studied classes with a large number of known members (classical
  novae,  Type Ia supernovae [SNe~Ia],  core-collapse  supernovae [CCSNe],  luminous  blue
  variables [LBVs]).  Vertically hatched  boxes show classes for which
  only few  ($\lessim4$) candidate members have been  suggested so far
  (luminous red novae,  tidal disruption flares, luminous supernovae).
  Horizontally hatched boxes are  classes which are believed to exist,
  but have not yet been  detected (orphan afterglows of short- and long
  GRBs).  The  positions of theoretically predicted  events (fall-back
  supernovae, macronovae, .Ia supernovae [.Ia]) are indicated by empty
  boxes.  See  text for references  related to each science  case. The
  brightest  transients  (on-axis   afterglows  of  GRBs)  and  events
  detectable predominantly  in the Milky  Way (e.g., dwarf  novae) are
  omitted for clarity (see Table~\ref{tab:transients_rates} for a more
  complete list).  Regions indicate the general location of each class
  and are not exclusive. The color of each box corresponds to the mean
  $g-r$  color  at  peak  (blue, $g-r<0$\,mag;  green,  $0\lessim  g-r
  \lessim 1$\,mag; red, $g-r>1$\,mag).}
\label{fig:magtau}
\end{figure}

The  Palomar Transient  Factory  (PTF) is  an  experiment designed  to
explore  systematically the  optical transient  and variable  sky. The
main  goal of  this project  is to  fill the  gaps in  our present-day
knowledge     of     the     optical     transient     phase     space
(Figure~\ref{fig:magtau}).      Besides     reasonably    well-studied
populations  (e.g.,  classical novae,  supernovae),  there exist  many
types of  either poorly constrained events (e.g.,  luminous red novae,
tidal disruption flares) or predicted but not yet discovered phenomena
(e.g.,  orphan afterglows  of $\gamma$-ray  bursts).   Here, dedicated
wide-field  instruments   have  the   best  prospect  of   leading  to
significant progress.

A number  of surveys (summarized in Table~\ref{tab:surveys}) have
attempted   this  challenging   task  and   have   provided  important
contributions  to  our  understanding  of  time-domain  science.   The
majority of these projects, however, have been designed (especially in
terms of  cadence) to maximize the discovery  probability for selected
source  populations  (typically   microlensing,  classical  novae,  or
supernovae).   Thus, large  areas  of the  phase  space remain  poorly
explored at best and are ripe for investigation with PTF.

The aim of this paper is to present the scientific goals of PTF and to
provide  predictions for discoveries  during the  first four  years of
operation. In order to give the  reader an overview of the project, we
start by summarizing the important instrumental aspects of PTF and the
general  survey strategy.  Much more  detailed discussion  of  the PTF
system,  both  hardware and  software,  can  be  found in  a  separate
publication \citep{law:2009}.

\begin{table*}[th]
\caption[]{Comparison of PTF with other untargeted transient and variable surveys}
\label{tab:surveys}
\begin{footnotesize}
 \begin{tabular}{lcccccccl}
\hline\hline
Survey & D\tablenotemark{a} & Scale & FoV & Cadence & m$_{R,{\rm lim}}$\tablenotemark{b} & Coverage & Lifetime & Reference\\
& [m] & [$''$ pxl$^{-1}$] & [deg$^{2}$] &  & [mag] & deg$^2$ night$^{-1}$ \\
\hline
Palomar Transient Factory & 1.26 & 1.0 & 7.78 & 1\,min -- 5\,d & 21.0 & 1000 & ongoing & Law et al. 2009\\
ROTSE-III\tablenotemark{c} & 0.45 & 3.25 & 3.42 & 1\,d& 18.5\tablenotemark{d} & 450 & ongoing & Quimby 2006\\
Palomar-Quest& 1.26 & 0.88 & 9.4 & 30\,min -- days & 21.0\tablenotemark{e} & 500 & 2003-2008 & Djorgovski et al. 2008\\
SDSS-II Supernova Search & 2.5 & 0.4 & 1.5 & 2\,d & 22.6 & 150 & 2005-2008 & Frieman et al. 2008a\\
Catalina Real-time Transient Survey& 0.7 & 2.5 & 8 & 10\,min -- yr & 19.5\tablenotemark{f} & 1200 & ongoing & Drake et al. 2008\\
Supernovae Legacy Survey & 3.6 & 0.08 & 1 & 3\,d-5\,yr & 24.3 & 2 & 2003-2008 & Astier et al. 2006\\
SkyMapper & 1.33 & 0.5 & 5.7 & 0.2\,d -- 1\,yr & 19.0 & 1000 & start 2009 & B. Schmidt, pers. com.\\
Pan-STARRS1 3$\pi$\tablenotemark{g}& 1.8 & 0.3 & 7 & 7\,d & 21.5 & 6000 & start 2009 & Young et al. 2008\\
Large Synoptic Survey Telescope & 8.4 & 0.19 & 9.62 & 3\,d & 24.5 & 3300 & start 2014 & Ivezic et al. 2008\\
\hline
\tablenotetext{a}{Telescope diameter}
\tablenotetext{b}{Typical limiting magnitude for single pointing}
\tablenotetext{c}{Texas Supernovae Search (till 2007) and ROTSE Supernovae Verification Project (since 2007)}
\tablenotetext{d}{Unfiltered}
\tablenotetext{e}{RG610 filter}
\tablenotetext{f}{$V$-band filter}
\tablenotetext{g}{Pan-STARRS1 3$\pi$ imaging survey.}
\end{tabular}
\end{footnotesize}
\end{table*}  

\section{Project Overview}
\label{sec:project}

\subsection{Instrument}
\label{sec:instrument}

PTF    uses   the    CFH12K   mosaic    camera   (formerly    at   the
Canada-France-Hawaii Telescope)  mounted on the  Samuel Oschin 48-inch
telescope  (P48)  at  Palomar  Observatory,  California.   The  camera
consists  of two  rows  of  six 2k$\times$4k  CCDs  each, providing  a
7.9\,deg$^{2}$ field  of view  (FoV) with a  plate scale  of 1.0\asec\
pixel$^{-1}$;  see \cite{rahmer:2008}  for more  details o the camera. Observations
will be performed mainly in  one of two broad-band filters (Mould-$R$,
sdss-$g$),    but   additional    narrow-band    filters   (H$\alpha$,
H$\alpha_{\rm off}$) can be  employed at the telescope.  Under typical
seeing  conditions (1.1\asec at  Palomar) the  camera achieves  a full
width  at half-maximum intensity  (FWHM) $\sim2.0$\asec  and 5$\sigma$
limiting  magnitudes   of  $R\approx21.0$,  $g^\prime\approx21.6$  and
H$\alpha\approx18$\,mag can be reached in a 60\,s exposure.

\subsection{Observing Strategy}
\label{sec:strategy}

The first few months after first light in December 13th 2008 have been
used  for  commissioning  of  the  instrument  and  software.  Initial
observations were taken to construct a fiducial reference image of the
sky for later use in transient detection.

Following  commissioning, PTF  will  be operating  on 80\%  ($\sim290$
night yr$^{-1}$)  of the P48  nights until at  least the end  of 2012.
During this time several predefined long-term surveys and a number of
evolving experiments  will be performed.  The  primary experiments are
the 5\,day cadence survey (5DC), the dynamic cadence experiment (DyC),
a    monitoring    of    the    Orion   star-forming    region    (see
\S~\ref{sec:orion}), and a narrow-band all-sky survey.  The respective
time allocations, cadences, and  photometric filters are summarized in
Table~\ref{tab:strategy}.

\begin{table}[th]
\caption[]{Summary of Main PTF Experiments}
\label{tab:strategy}
\begin{center}
 \begin{tabular}{lccc}
\hline\hline
Experiment & Exposure & Cadences & Filter \\
& \% of total &  &  \\
\hline
5DC  & 41 & 5\,d & $R$ \\
DyC &  40 & 1\,min -- 3\,d & $g,R$\\
Orion &  11 &  1\,min & $R$ \\
Full Moon  & 8 & -- & H$\alpha$\\
\hline
\end{tabular}
\end{center}
\end{table}  

The 5DC  experiment will at any  given time monitor an  active area of
$\sim2700$\,deg$^2$ with  a mean  cadence of 5\,d.   Over the  year, a
total footprint  of $\sim10,000$\,deg$^{2}$ (largely  overlapping with
the    SDSS)    with     Galactic    latitude    $|b|>30$\degs,    and
$-15$\degs$<\delta_{\rm   J2000}<86$\degs  will  be   observed.   This
experiment will be performed  using the $R$-band filter, and typically
two 60\,s  exposures separated by 1\,hr  will be taken  in each epoch.
As part  of the 5DC, we  will also observe  $\sim60$ fields containing
nearby galaxies.  Note that these galaxies (with the exception of M31)
will cover only a small fraction of the camera FoV.

The  DyC experiment  is  designed to  explore  transient phenomena  on
time scales  shorter than  5\,d and  longer than  about  1\,min.
Here,  several  different  $g$-   and/or  $R$-band  surveys  will  be
conducted  during the first  six months  of operation.  Strategies for
subsequent DyC observations will  be decided after the initial results
have been analyzed.



\begin{figure}
\begin{center}
\centering
\includegraphics[width=0.48\textwidth]{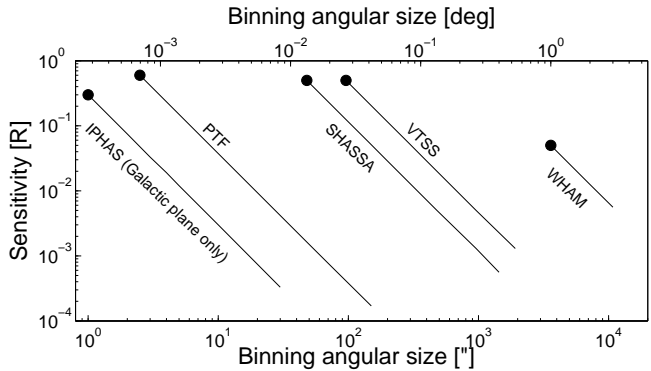}
\end{center}
\caption{The  surface  brightness  sensitivity  (Rayleigh, R)  of  the
  PTF  H$\alpha$ survey  as a function  of the  angular size.
  The smallest  angular size is  determined by the pixel  size (filled
  circle). Binning  is expected to improve sensitivity  in proportion to the square root
  of  the number  of  pixels.  The proposed  survey  will have  better
  sensitivity  by more  than  two orders  of  magnitude over  previous
  surveys     (e.g.,     SHASSA;     Gaustad et al. 2001,     VTSS;
  Dennison et al. 1998). The only  survey with superior sensitivity
  \citep[IPHAS;][]{walton:2004}  is  limited  to the  Galactic  plane
  ($|$b$|<5$\degs)}
\label{fig:halpha}
\end{figure}

During  bright time  $3\pi$\,sr narrow-band  (H$\alpha$, H$\alpha_{\rm
  off}$, [O~III])  surveys will  be conducted. Here,  $\sim4000$ tiles
covering the sky at  $\delta_{\rm J2000}>-28$\,\degs\ will be observed
at least  twice in each  filter. The estimated $5\sigma$  point source
flux     limit     for     the     H$\alpha$    survey     will     be
$\sim2\times10^{-17}$\,erg\,cm$^{-2}$\,s$^{-1}$ ($\sim0.6$\,Rayleigh).
Better sensitivity by a factor  of $\sim$60 can be achieved by binning
the H$\alpha$ images  to a resolution of $1'$.   This sensitivity will
constitute an improvement  over that of previous surveys  by a factor
of $\sim250$ (Figure~\ref{fig:halpha}).

\subsection{Follow-up Strategies}
\label{sec:followup}

The key to  a successful transient survey lies  in the availability of
follow-up    resources,   specifically    multi-color    imaging   and
spectroscopy. Most other wide-field  searches aim to do both discovery
and  follow-up  observations,  with   the  same  instrument,  or  have
spectroscopic follow-up biased heavily  toward a specific science goal
(e.g., ESSENCE, SNLS, and SDSS-II with SNe~Ia).  In contrast, PTF will
perform multi-color  photometry for  all candidate transients  using a
variety  of facilities.   Here, a  large fraction  of  Palomar 60-inch
\citep[optical;][]{cenko:2006},       LCOGT\footnote{http://lcogt.net/}
(optical)      Super-LOTIS     \citep[optical;][]{park:1999},     KAIT
\citep[optical;][]{filippenko:2001},            and           PAIRITEL
\citep[near-IR;][]{bloom:2006}  time has  been set  aside.  Additional
spectroscopy is planned  at the Palomar Hale 5-m,  MDM Observatory, William Herrschel 4.2~m, and
Lick 3~m telescopes.

A detailed discussion of all aspects mentioned in this section as well
as  the  data reduction,  analysis,  and  archiving  can be  found  in
\cite{law:2009}.

\section{PTF Science Goals}
\label{sec:science}
The PTF design is optimized for  repeated coverage of a large sky area
in a short  amount of time.  This is crucial  for pursuing the primary
science  goal,  namely  to  investigate  the  various  populations  of
transients and variables.  In addition, it allows the usage of deeper,
coadded  images to  study  persistent sources,  ranging from  Galactic
foreground stars  to distant  galaxies.  While it  is not  possible to
predict all  of the science that PTF  will enable, a few  key areas of
interest have  been defined  and will be  summarized below.   Here, we
first address transient and  variable sources from our neighborhood to
high redshift, followed by regularly variable classes and science with
the coadded  data products.  An overview  of the sections  is given in
Table~\ref{tab:sections},    while    Table~\ref{tab:transients_rates}
summarizes the  properties, universal  rates, and PTF  predictions for
transients and variables.

\begin{table}[th]
\caption[]{Summary of Science Drivers discussed in \S~\ref{sec:science}}
\label{tab:sections}
\begin{center}
 \begin{tabular}{lcc}
\hline\hline
Science Driver & Main Survey & Section \\
\hline
Cataclysmic variables & DyC, 5DC & \ref{sec:cv}\\
Classical novae & 5DC & \ref{sec:cn}\\
Luminous red novae, .Ia supernovae & DyC, 5DC & \ref{sec:other_nearby}\\
Type Ia supernovae & 5DC & \ref{sec:snIa} \\
Core-collapse supernovae & 5DC & \ref{sec:ccsn}\\
Luminous \& pair-instability supernovae & 5DC & \ref{sec:pisn}\\
Blazars & 5DC & \ref{sec:blazars}\\
Tidal disruption flares & 5DC & \ref{sec:tdf}\\
(Orphan) GRB afterglows & DyC, 5DC & \ref{sec:afterglows}\\
Exoplanet transits in Orion & Orion & \ref{sec:orion}\\
Eclipsing Objects Around Cool Stars& DyC & \ref{sec:mdwarfs}\\
RR~Lyrae stars& 5DC & \ref{sec:rrlyrae}\\
Galactic Variable Stars & DyC,5DC & \ref{sec:otherVars}\\
Microlensing & 5DC & \ref{sec:micro}\\
Near-Earth objects & 5DC & \ref{sec:neo}\\
PTF as follow-up instrument& ToO & \ref{sec:too}\\
Coadded \& narrow- band data & Dyc, 5DC, NB & \ref{sec:surveys}\\
\hline
\end{tabular}
\end{center}
\end{table}  


\subsection{Cataclysmic Variables}
\label{sec:cv}

Cataclysmic variables (CVs) form a broad family of highly variable and
dynamical  stellar binaries  in  which matter  from  a low-mass  donor
(M$\lessim 1$\,\msun) is accreted  onto a white dwarf \citep[M$\approx
1$\,\msun;  e.g.,][]{warner:1995}.  Variability, from  milliseconds to
hundreds  of  years, arises  from  different  physical processes,  and
conclusions  derived  from  CVs  have  been  extrapolated,  upward  or
downward in scale,  to other phenomena such as  active galactic nuclei
(AGNs) and X-ray binaries.  Despite being studied for several decades,
many details concerning the  viscous turbulence in the accretion disk,
interaction of  the accreted matter  with the atmosphere of  the white
dwarf,  irregularities within  regular photometric  behavior,  and the
final fate of these objects, are still missing.

An important subclass of CVs is the so-called dwarf novae (DNe), which
show quasi-periodic outbursts (on time scales  of weeks to years) with
amplitudes of  3--8\,mag and  durations of typically  3--20\,d.  These
transient events  are the result of  a sudden viscosity  change in the
accretion  disk.  The origin  of this  change is  still controversial;
instabilities in the disk  \citep{meyer:1981} or in the secondary star
\citep{bath:1986} may  both be  responsible. Increasing the  sample of
well-studied events will help to distinguish between these models.

The  current  estimate  of  the  space  density  of  DNe  ranges  from
$\sim3\times10^{-5}$\,pc$^{-3}$   \citep[observations;][]{schwope:2002}
to  as much as  $\sim10^{-4}$\,pc$^{-3}$ \citep[theory;][]{kolb:1993}.
For typical  values of peak magnitude  ($9>M_R>4$\,mag) and recurrence
time scale \citep[1\,yr;][]{rau:2007b}, this suggests the detection of
about 100 DNe yr$^{-1}$ in the 5DC.

In  special cases,  the  secondary can  be  another white  dwarf or  a
hydrogen-depleted star  in a tight orbit, whose  evolution is dictated
by  gravitational wave  radiation.  Depending  on the  orbital period,
mass transfer can  either be through a stable or  unstable disk or via
direct impact. These so-called AM~CVn stars are of particular interest
as  they form the  dominant population  of gravitational  wave sources
detectable with  the future Laser Interferometer  Space Antenna (LISA)
mission. Merging  white dwarfs  are also strong  progenitor candidates
for   SNe~Ia  \citep[e.g.,][]{iben:1984,webbink:1984}  and   thus  are
important tracers of binary stellar evolution.

AM~CVn  systems with  an unstable  accretion disk  exhibit photometric
outbursts similar  to those  of hydrogen-rich dwarf  novae and  can be
detected  as transient  events with  PTF.  However,  with  an inferred
space                            density                            of
$\sim(1-3)\times10^{-6}$\,pc$^{-3}$\citep{roelofs:2007},  AM~CVn stars
are rare  compared with  normal CVs.  Furthermore,  only 5-6\%  of the
population  resides within  the instability  range in  which outbursts
occur, and He ionization results  in much shorter time scales of these
events. Thus, only  a few of these elusive sources  are expected to be
found in the DyC experiments.


\subsection{Classical Novae}
\label{sec:cn}

In  some CVs  the matter  accreted onto  the primary  white  dwarf can
experience a thermonuclear runaway, driving substantial mass loss from
the  system.   These  classical  nova eruptions  reach  absolute  peak
magnitudes   of   $-5\gtrsim   M_R\gtrsim-10$\,mag   ($<M_R>\sim-7.5$\,mag in  M31; Shafter et al. 2009) and  fade on time
scales of days to weeks \citep[e.g.,][]{warner:2008}.

A  typical   M31-like  galaxy  produces   $\sim2\times10^{-10}$  novae
yr$^{-1}$ L$^{-1}_{\odot,K}$  \citep{ferrarese:2003}, translating into
$\sim$30  yr$^{-1}$.   PTF  will  repeatedly image  60  large,  nearby
(m-M\,$<29$\,mag) galaxies as part of the 5DC.  We anticipate $\sim$10
novae yr$^{-1}$ galaxy$^{-1}$ to  lie above the detection limit. Thus,
a  typical coverage  of three  months per  year for  each  galaxy will
provide  $\sim$150  extragalactic  classical nova  discoveries  every
year.

The spatial  and luminosity distributions of these  novae inside their
host  galaxies  will provide  invaluable  tests of  cataclysmic-binary
evolution  theory.  Predictions that  fast novae  should predominantly
arise  in  spiral   arms  \citep{dellaValle:1994}  are  still  lacking
conclusive  observational  evidence \citep[e.g.,][]{neill:2004}.   The
expected dramatic enlargement of the number of accurately sampled nova
light curves will help to test this prediction.

In addition  to targeting  nearby galaxies, PTF  will place  the first
meaningful  limits  on  the  space density  of  intergalactic  ``tramp
stars'' \citep{shara:2006}  by searching for classical  novae as their
proxies.  Previous detections of  ``tramp" planetary nebulae and novae
in  the Virgo  cluster \citep{feldmeier:2004}  and the  Fornax cluster
\citep{neill:2005}  suggest that  16\% to  40\% of  all stars  in rich
clusters reside  in intracluster space outside of  galaxies.  PTF will
probe whether  the same  is true of  less dense environments  like the
Local Group, or of intergalactic  space in general.  In the 5DC, novae
originating  in a space  volume at  least four  times larger  than the
Local Group  can be detected (out  to a distance  of $D=2.5$\,Mpc). We
expect $\sim$16  tramp novae  to be  found every year  if 10\%  of the
stars in  and around the Local Group  are tramps\footnote{This assumes
  8\,months per  year observation of the  $\sim$2700\,deg$^{2}$ of the
  5DC and $\sim$60 novae yr$^{-1}$ detected in the Local Group (mostly
  M31 and the Galaxy).}.  Conversely, if zero intergalactic tramps are
found, an upper limit on  the baryon contribution in the Universe from
intergalactic tramp stars of $\sim$1--2\% can be set after one year.

\subsection{Other Transients in Nearby Galaxies}
\label{sec:other_nearby}

Observations suggest a paucity of transients between the peak absolute
brightness of classical  novae ($M_{R}\gtrsim-10$\,mag) and supernovae
($M_{R}\lessim-14$\,mag).   However, recent discovery  of a  number of
enigmatic  events in  this  gap have  demonstrated  the potential  for
exciting discoveries with dedicated experiments.

One of these newly emerging classes are the luminous red novae (LRNe),
whose origin and explosion  physics still pose unsolved puzzles.  This
elusive  group of  sources includes  only four  known  members: M31~RV
\citep{rich:1989},     V4332~Sgr     \citep{martini:1999},    V838~Mon
\citep{brown:2002},  and M85~OT\,2006-1  \citep{kulkarni:2007}.  Their
observational  properties differ  from  those of  classical novae  and
typical  supernovae by  showing  a slowly  evolving  outburst with  an
optical plateau lasting weeks to  months, followed by a strong redward
evolution resembling a transition from stellar type $\sim$F to $\sim$M
and colder \citep[e.g.,][]{rau:2007a}.

Several formation models have been proposed, including stellar mergers
\citep{soker:2003},  rare   novae  \citep{iben:1992},  and  supernovae
\citep{pastorello:2007}; these  scenarios need to be  assessed with an
increased sample  of accurately studied events.   Also, the underlying
stellar population of the known  sample has a large diversity, ranging
from a B-star cluster in the Milky Way \citep[V838~Mon;][]{afsar:2007}
to the  bulge of M31 \citep[M31~RV;][]{rich:1989}.  This  leads to the
question  of whether  the known  LRNe  form one  homogeneous class  of
objects (with  the explosion physics being independent  of the stellar
population), or if different scenarios apply to each of the individual
sample members. Note that two  additional sources at the bright end of
the  distribution,  SN~2008S \citep[e.g.,][]{arbour:2008,prieto:2008b}
and NGC300~OT \citep[e.g.,][]{monard:2008},  have been found recently.
Associations  with extreme  asymptotic giant  branch (AGB)  stars have
been proposed for these events \citep{thompson:2008}.

Based   on   rates  estimated   from   the   number   of  known   LRNe
\citep[1.5$\times10^{-13}$\,yr$^{-1}$
L$_{\sun,K}^{-1}$;][]{ofek:2008},     we     expect    to     discover
$>1.5$\,yr$^{-1}$   during  the  5DC   coverage  of   galaxies  within
16\,Mpc. Over a four-year lifetime  of PTF, this will more than double
the  known sample  and provide  valuable insight  into the  origin and
properties of these rare transients.

The regular PTF monitoring of nearby galaxies will also allow a search
for a  number of theoretically predicted, but  not yet observationally
confirmed,  populations  of transients.   Among  these  are the  faint
thermonuclear  supernovae  from  AM~CVn binaries  \citep[``.Ia  SNe'',
one-tenth    as    bright    for    one-tenth   the    time    as    a
SN~Ia;][]{bildsten:2007}.  Helium  that accretes onto  the white dwarf
primary  can undergo  unstable thermonuclear  flashes. At  wide binary
separations (P$_{\rm  orb}>25$\,min) a large mass  will be accumulated
before  ignition, leading to  violent flashes  in which  high pressure
allows the  burning to produce  radioactive elements to power  a faint
($-15\gtrsim  M_{  R}\gtrsim-17$\,mag)  and rapidly  rising  (2--5\,d)
thermonuclear  supernova.   The local  Galactic  AM~CVn space  density
implies one such explosion every 5,000--15,000\,yr in $10^{11}M_\odot$
of  old stars  ($\sim$2--6\% of  the  SNe~Ia rate  in E/SO  galaxies),
giving an  expected DyC yield  of a few  per year. The  first possible
manifestations of this  class in nature have recently  been found with
SN~2008ha \citep{foley:2009} and SN~2005E \citep{perets:2009}.



\subsection{Type Ia Supernovae}
\label{sec:snIa}

Observations of  SNe~Ia provided  the first direct  and, to  date, the
best evidence for  the acceleration of the expansion  of the Universe,
propelled    by   some    kind   of    mysterious    ``dark   energy''
\citep[][]{riess:1998,perlmutter:1999}; see Frieman et al. (2008b) for
a review. To determine  whether the equation-of-state parameter of the
dark  energy\footnote{$w =  P/\rho$ with  $P$ being  the  pressure and
  $\rho$  is  the  energy  density.}  is  consistent  with  Einstein's
cosmological constant,  enormous efforts are underway  to populate the
SN~Ia Hubble diagram with  accurate measurements of hundreds of events
at redshifts $0.1 < z < 1.0$. Most noteworthy are the Supernova Legacy
Survey  \citep[SNLS;][]{astier:2006},  ESSENCE \citep{miknaitis:2007},
and  the  SDSS  SN  survey  \citep{frieman:2008a}.   Future  precision
cosmology experiments  (e.g., Joint Dark  Energy Mission) may  work at
even higher redshifts (up to $z \approx 1.7$).

SN~Ia cosmology  relies on comparison of the  apparent brightnesses of
high-redshift  events to those  at low  redshift, to  measure accurate
relative distances.   PTF is expected to  discover $\approx500$ SNe~Ia
yr$^{-1}$  during the  5DC in  the  nearby smooth  Hubble flow  ($0.03
\lessim z \lessim  0.14$) out of which $\approx150$  yr$^{-1}$ will be
followed in detail, all caught  well before maximum light and  in the lower half
of this redshift range. 

The  discrimination among specific  models of  dark energy  requires a
more  detailed understanding  of SNe~Ia  and  on methods  to test  for
cosmic  evolution and  other systematic  effects, e.g.,  separation of
intrinsic SN color-luminosity relations from dust reddening, impact of
the environment, and evolution  of the progenitors with redshift.  See
Howell {et al.} (2009)  for  a  review  of  the  largest  systematic
uncertainties affecting  SN Ia and methods to  mitigate their effects.
Owing to the augmented low-redshift statistics and systematics control
of  the program,  the PTF  light  curves, in  conjunction with  higher
-redshift data, will allow the construction of the most accurate SN~Ia
Hubble diagram yet.

\begin{figure}
\begin{center}
\vspace{-0.01in}
\centering
\includegraphics[width=0.45\textwidth,angle=0]{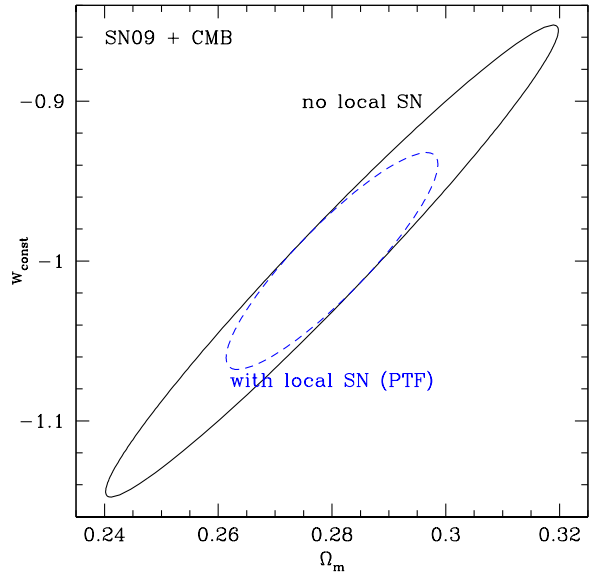}
\end{center}
\caption{ Impact  of 1000 well-studied,  well-calibrated, low-redshift
  Hubble-flow SNe~Ia  when mated to  near-term simulated high-redshift
  surveys ($z=0.9$).  }
\label{fig:snap}
\end{figure}

The purely statistical impact of a large number of low-redshift SNe~Ia
on  the  determination of  the  cosmic  parameters  is illustrated  in
Figure~\ref{fig:snap}. A simulated data set of 1000 SNe~Ia centered at
$z=0.08$ is paired  with 500 SNe~Ia distributed in  the redshift range
$0.2<z<0.8$ The constraints  on a constant form of  the dark energy as
well as on the dynamical component are strengthened by a factor of two
when  the local  sample  is included.   These  results illustrate  the
impact that the PTF data set can have in the near future. In addition,
current SN  Ia analyses are dominated by  systematic uncertainties and
the  homogeous,  well-calibrated  sampling  of  PTF  will  provide  an
important improvement.

The large  area covered by the  5DC experiment will  include a sizable
number  of  relatively  nearby,  Abell-like galaxy  clusters.   SNe~Ia
occurring in  these clusters will  be of particular interest.   The SN
rate  in  galaxy  clusters  is   a  useful  tool  to  constrain  SN~Ia
progenitors \citep[e.g.,][]{maoz:2004,sharon:2007}, while the fraction
of  hostless  SNe   \citep{gal-yam:2003},  presumably  resulting  from
progenitors that  are members of the  intergalactic stellar population
in  clusters, is  a  useful  probe of  the  fraction of  intergalactic
``tramp'' stars and its evolution with redshift.

PTF's capability  to reveal a large  number of SN~Ia  has already been
demonstrated by the first discoveries during commissioning. Two events
were found by  searching 589\,deg$^2$ of data taken  in a single night
on  March  2nd  2009  \citep{kulkarni:2009}.   The  first  source  was
detected at $18.62\pm0.04$\,mag  near the core of an  SDSS galaxy with
g=17.78 and a redshift of z=0.0555 and spectroscopically identified as
SN~Ia  at the  Palomar Hale  5m telescope.   The second  event  was an
independent  detection   of  SN  2009an   \citep{cortini:2009}.   More
information about the first PTF  transient discoveries can be found in
\cite{law:2009}.

\subsection{Core-Collapse Supernovae}
\label{sec:ccsn}

The  study   of  core-collapse  supernovae  (CCSNe)   is  critical  to
understanding  the  final stages  of  massive-star  evolution and  the
formation of  neutron stars and black holes,  natural laboratories for
general relativity.  Furthermore,  CCSNe produce heavy elements, dust,
and  cosmic rays; their  explosion shocks  trigger (and  inhibit) star
formation;  and their  energy input  to the  interstellar medium  is a
crucial ingredient in modeling galaxy formation.

CCSNe  will  likely  form  one   of  the  most  luminous  and  distant
populations  of  transients  for  PTF.  Their  typical  peak  absolute
magnitudes\footnote{Determined      from       72      CCSNe      from
  \cite{richardson:2002}  assuming  M$_B-$M$_R\approx0$\,mag.}   range
from   $M_{R}=-14$  to  $-21$\,mag   ($<M_R>\approx-17.8$\,mag),  thus
allowing detections out to $z\approx0.4$ for the most luminous events.
Based  on  discovery statistics  from  the  Supernova Factory  project
\citep{aldering:2002}, the 5DC is expected to reveal $\sim200$\,events
yr$^{-1}$ shortly after explosion down to $R\approx19.5$\,mag.

\begin{figure}
\begin{center}
\vspace{-0.01in}
\centering
\includegraphics[width=8.5cm,angle=0]{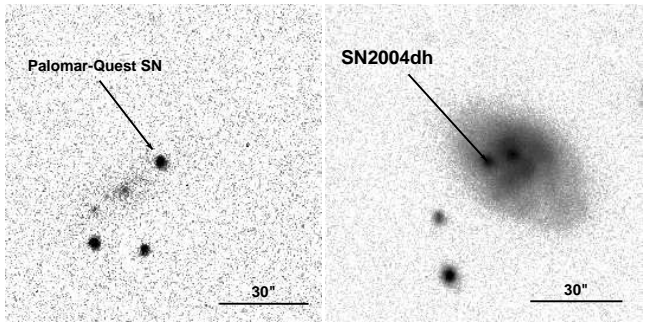}
\end{center}
\caption{Comparison  of two  CCSNe at  similar redshift  and magnitude
  from  the  Caltech  Core-Collapse Program  \citep[][]{gal-yam:2007}.
  The  left and  right  images show  an  unnamed SN  discovered in  an
  untargeted \citep[SN  Factory;][]{aldering:2002} and SN~2004dh found
  in   a    targeted   search   \citep[Lick    Observatory   Supernova
  Search;][]{li:2000}), respectively.  Note  the LMC-like faintness of
  the host in the left  panel compared to the luminous SN~2004dh host.
  Such low-luminosity (and  probably low-metallicity) hosts are likely
  to  produce  new  types  of  SNe.   Thus, a  sample  drawn  from  an
  untargeted survey such  as the PTF 5DC will  provide a more accurate
  picture of the cosmic SN population.  }
\label{fig:cccp}
\end{figure}

Most  nearby SNe  are  discovered by  repeated  imaging of  catalogued
galaxies \citep[e.g.,][]{filippenko:2001}.  This introduces a possible
bias, skewing  the resulting sample  strongly toward events  in large,
luminous,   metal-rich  galaxies   (Figure~\ref{fig:cccp}).   Previous
untargeted surveys  have uncovered  peculiar SNe, occurring  in small,
low-luminosity,       and      probably       metal-poor      galaxies
\citep[e.g.][]{young:2008}.   Metallicity has a  strong impact  on the
pre-SN  evolution of  massive stars,  influencing wind  mass loss (and
thus the final  core mass and composition), loss  of angular momentum,
and  the  opacity  \citep[e.g.,][]{fryer:2007}.   Also,  long-duration
gamma-ray bursts  (GRBs), which are associated  with very energetic
CCSNe,  have  been  found  in  faint,  low-metallicity  host  galaxies
\citep[e.g.,][]{stanek:2006,modjaz:2008}.

The  5DC provides  an untargeted  search with  thousands  of anonymous
galaxies in each  frame, allowing one to construct  a supernova sample
without host-galaxy  bias.  This will  provide an accurate  picture of
the cosmic  CCSN population  and present observational  constraints on
theoretical models for low- and high-metallicity explosions.  Assuming
galactic luminosity functions measured by  the SDSS at $z=0.1$ and the
linear  metallicity-luminosity relation \citep{tremonti:2004},  we can
estimate  that at least  $\sim15$\% of  the SNe  detected by  PTF will
reside in  galaxies which  are as metal  poor as the  Large Magellanic
Cloud (LMC),  and about $20$\% of  those will reside  in galaxies more
metal-poor than the Small Magellanic Cloud (SMC).

SNe~II-P, a subset of CCSNe,  have been shown to be standardizable by
an   empirical   correlation  \citep{hamuy:2002,nugent:2006}.    Their
intrinsic  inhomogeneity  can be  calibrated  well  and the  resulting
distance precision for cosmology is about 10\% \citep{poznanski:2009},
only      slightly      worse       than      that      of      SNe~Ia
\citep[7\,\%;][]{astier:2006}. PTF  will populate the  SNe~II-P Hubble
diagram  with  about 20  Hubble-flow  events  per  month to  build  an
entirely  independent and novel  test of  the cosmology  inferred from
SNe~Ia.

Finally,  with  dense  sampling   of  virtually  every  nearby  galaxy
observable  from  the northern  hemisphere,  PTF  will  be a  powerful
discovery  machine for  the most  under-luminous (or  rapidly decaying)
events, providing additional clues to the nature of cosmic explosions
with           very          faint           optical          displays
\citep[e.g.,][]{gal-yam:2006b,bildsten:2007,pastorello:2004}.

\subsection{Luminous Supernovae and Pair-Instability Supernovae}
\label{sec:pisn}

The first  stars to form in  the Universe typically  ended their brief
lives           through           pair-instability          supernovae
\citep[e.g.,][]{barkat:1967,heger:2002}, but it was long believed that
this exit was not available  to the metal-enriched stars of the modern
era.  Recently, however, the  discovery of SN~2006gy has provided what
may    be    the    first    evidence   for    such    an    explosion
\citep{ofek:2007,smith:2007},   and   surprisingly,   this   extremely
luminous (brighter  than $-20$\,mag for  150\,d), long-lived supernova
was  found  in the  local  Universe.   The pair-instability  mechanism
requires a star to meet its end with a substantial mass still bound to
it.  The  rate  of  such  events  in  the  local  Universe  will  thus
significantly augment our understanding of  the top end of the stellar
initial  mass  function as  well  as  the  role metallicity  plays  in
shedding mass  prior to the  explosion.  As SN~2006gy-like  events are
luminous,  expected to  occur frequently  in the  early  Universe, and
relatively simple to  model, they may serve as  cosmological probes in
the James Webb  Space Telescope era.  Events discovered  by PTF in the
local Universe may serve as a proving ground for their utility.


The  estimated event rate  of $\sim$100\,Gpc$^{-3}$  yr$^{-1}$ suggest
that about 20 (200)  SN~2006gy-like events yr$^{-1}$ above $R=18$\,mag
(21.0\,mag) can  be discovered  during the course  of the 5DC.   It is
important  to note that  the handfull  of known  events have  all been
discovered   in   blind  surveys   (e.g.,   Texas  Supernova   Search,
\citep{quimby:2006};    Catalina     Real-time    Transient    Survey,
\citep{drake:2009}), such as PTF.

The most luminous supernova yet  identified is SN~2005ap, found by the
TSS \citep{quimby:2007}.   Smith et al. (2008)  suggest that SN~2005ap
may be physically similar to  SN~2006gy, and Quimby et al. (2007) note
a possible connection to the  engines powering GRBs.  With only weakly
diluted   black-body   spectra   and  rapid   photometric   evolution,
SN~2005ap-like  events  offer  another  possible probe  of  the  early
U1niverse, and  PTF can help  reveal their physical identity.  While a
rate cannot accurately  be measured from a single  event, we speculate
that  SN~2005ap-like events  are  roughly half  as  common as  sources
similar to SN~2006gy.

While  there  is  not  yet  any conclusive  evidence  that  SN~2005ap,
SN~2006gy, or any of the  extremely luminous SNe are indeed related to
Pair-Instability explosions, there is no conclusive evidence that they
are  not.  PTF will  significantly expand  the sample  of well-studied
luminous SNe and perhaps provide an  answer as to whether or not PISNe
occur in the local Universe.

\subsection{Blazars}
\label{sec:blazars}

The light  curves of blazars  \citep[AGN with the jet  pointed towards
the  observer; e.g.,][]{urry:1995}  are  dominated by  Doppler-boosted
instabilities  and   other  phenomena  associated   with  relativistic
outflows, rather than by  the accretion-driven variability of ordinary
AGN.  Thus, variability is one of their most notable characteristics,
offering both a method for discovery and insight into their physics.

Blazars   are  likely   the  dominant   contributors  to   the  cosmic
$\gamma$-ray background  \citep{narumoto:2006,giommi:2006}.  They also
form the main foreground population  at high radio frequencies, and an
understanding of  them will  be essential for  proper modeling  of the
cosmic  microwave background  radiation at  high  angular frequencies,
such      as     for      the      Planck     mission      \citep[see,
e.g.,][]{giommi:2006,giommi:2007,wright:2008}.   Moreover,  a  growing
number   of    blazars   has    been   detected   at    TeV   energies
\citep[e.g.,][]{horan:2004,horns:2008}. Even more intriguingly, beamed
AGN  are perhaps  among the  most plausible  sources  of ultra-high  energy
cosmic     rays,     reaching     energies     of     $10^{21}$     eV
\citep[e.g.,][]{dermer:2007b}.  These cosmic accelerators are bound to
play an  increasingly important role  in the future  of astro-particle
physics.



The {\it  Fermi} mission will play  a dominant role in  the studies of
blazars  over  the  next  several years.   Ground-based  studies  from
synoptic sky surveys like PTF will provide valuable complementary data
for   maximum  scientific  returns.    First,  archival   analysis  of
multi-epoch data  from PTF, combined with other,  previous and ongoing
synoptic surveys such as  Palomar Quest \citep{djorgovski:2008} or the
Catalina Sky  Survey \citep{drake:2009}, can be used  to define purely
variability-based  (or  variability-   and  color-based)  samples  of
blazars. This will provide an important check on the selection effects
in the  more traditional, radio-based samples.   Archival light curves
and measures of variability can be  used in a joint analysis with data
from  radio  to  $\gamma$-rays.   This  will result  in  larger,  more
complete samples of blazars, crucial for the studies described above.

Second, real-time  detection of optical flares  of blazars, correlated
with the  $\gamma$-ray data  stream from {\it  Fermi}, can be  used to
detect  $\gamma$-ray loud  blazars at  fainter flux  levels  than what
would be statistically justifiable from the $\gamma$-rays alone, or to
lead to new blazar discoveries  on their own.  These detections can be
used to trigger other follow-up  observations; a joint analysis of
such  multi-wavelength, synoptic observations  of blazar  flares could
lead to  better constraints for theoretical  models.  Such statistical
monitoring  of large areas  of the  sky is  complementary to  the more
traditional follow-up monitoring of  selected samples of known blazars
\citep[e.g.,][]{carini:2004}.  We estimate that  PTF will see at least
a few tens of potential {\it  Fermi} blazars per clear night, of which
roughly one will have a strong optical blazar flare.

Blazars are known to vary  significantly on all time scales where data
exist, with the  power spectra well represented by a  power law, so in
principle  an   arbitrarily  large  fluctuation  can   occur  given  a
sufficiently long time interval.  In practice, fluctuations at a level
of several tenths of a magnitude are common on time scales of hours or
days, and fluctuations at a level  of 1 or 2\,mag are commonly seen on
time scales of months,  but with rise or fall times as  short as a day
or  even less.   This should  be  contrasted with  the variability  of
normal quasars, where typical variations  on a scale of several tenths
of a  magnitude occur on  a time scale  of years, and  dramatic (e.g.,
magnitude or  greater) changes are  extremely rare on the  time scales
probed by the existing observations.

\subsection{Tidal Disruption Flares}
\label{sec:tdf}

Stars passing within a  distance of $5 M_7^{-2/3}$ Schwarzschild radii
of a supermassive black  hole (SMBH) of $M_{\rm SMBH}=10^7M_7$\,\msun\
will  be   torn  apart  by   the  strong  tidal   gravitational  field
\citep[e.g.,][]{hills:1975,rees:1988}.   While for  $M_7\gtrsim20$ the
disruption of  a main-sequence star happens inside  the event horizon,
for less massive SMBHs this should  give rise to a detectable flare of
emission.   To date,  only  a few  candidate  tidal disruption  flares
(TDFs) have  been found, primarily in  the {\it ROSAT}  All Sky Survey
archive \citep[e.g.,][]{li:2002,gezari:2003,halpern:2004,komossa:2005}
and   in  restframe-UV  monitoring   of  otherwise   dormant  galaxies
\citep[e.g.,][]{renzini:1995,gezari:2006,gezari:2008b}.

The flare emission is predicted  to be dominated by an optically thick
accretion disk with  a temperature $T_{\rm eff} \approx  3 \times 10^5
M_7^{-1/4}$\,K  \citep{ulmer:1999},  peaking   in  the  far-UV.   Line
emission  from  unbound  (ejected)  material  \citep{strubbe:2009}  is
expected to significantly enhance the black-body continuum flux in the
optical.  This can lead to  a peak absolute magnitude in the rest frame
of $M_{R}  \approx -15$  to $-19$\,mag (see  also Strubbe  \& Quataert
2008) and  observationally confirmed by Gezari et  al. 2008), roughly
comparable to the absolute magnitude of a supernova.  The single-epoch
PTF sensitivities  suggest that the  TDF population will be  probed to
$\sim 200 M_7$\,Mpc ($z=0.05$),  corresponding to a volumetric rate of
$40 M_7^{3/2}$ \,yr$^{-1}$ over the entire sky or about 3 yr$^{-1}$ in
the 5DC.


One  of the  principal difficulties  in  pursuing these  events is  in
distinguishing    TDFs   from   nuclear    SNe   and    ordinary   AGN
variability. Here both  the particular light curve (a  rapid rise over
days to weeks followed  by a power-law decline\footnote{The decline of
  interloping  SNe  should be  exponential.})  and spectral  evolution
provide  valuable guidance.   Historical variability  limits  from the
DeepSky
project\footnote{http://supernova.lbl.gov/$\sim$nugent/deepsky.html .}
can further constrain patterns of low-level AGN activity.

We expect roughly  10\% of rapidly risig events  in galactic nuclei to
be  TDFs  (as opposed  to  SNe).   Having  $\sim$2\,kpc resolution  at
$z=0.05$, we must  contend not only with nuclear  transient events but
the  light  of  the  host-galaxy  bulge  itself.   Thus,  events  from
lower-mass  ($10^5$--$10^6$\, M$_\odot$)  SMBHs may  be  more reliably
identified since their bulges are expected to be comparatively faint.

Even if the yield from PTF  is only a few iron-clad cases, the ensuing
multi-frequency  investigations  will  have  great value  in  providing
feedback for TDF models and  for influencing strategies for future TDF
observing campaigns.   These discoveries will provide a  window into the
demographics of SMBHs and  detailed follow-up observations could allow
for an entirely new way of measuring black hole mass, thus offering an
independent  test  of   the  $M_{\rm  SMBH}$--$\sigma_*$  relation  of
galaxies \citep[e.g.,][]{ferrarese:2000,gebhardt:2000,tremaine:2002}.
 
\subsection{Untriggered and Orphan GRB Afterglows}
\label{sec:afterglows}

By far the most luminous  (up to $M_R=-37$\,mag) transients that
PTF   may  detect   are  the   afterglows  of   GRBs  for   which  the
ultrarelativistic  collimated outflows are  pointed toward  the Earth.
GRB statistics from  BATSE onboard {\it CGRO} suggest  that a few such
events  happen  every   day  in  the  Universe  \citep{paciesas:1999},
corresponding  to  $\sim0.02$\,deg$^{-2}$  yr$^{-1}$.  Out  of  these,
$\sim50$\% remain  brighter than $R=20$\,mag for at  least 1000\,s and
can in  principle be discovered blindly\footnote{By this  we mean, not
  by   follow-up  observations in  response   to   a   trigger  with   high-energy
  observations.}  with  PTF.  These untriggered  afterglows may appear
as fast transients in a single  or pair of images of the 5DC; however,
the expected rate is very small ($\sim0.5$\,yr$^{-1}$).

A  more  numerous  population   of  transients  is  predicted  by  the
collimated  nature of  GRBs,  observationally suggested  by the  ``jet
breaks''          in          afterglow          light          curves
\citep[e.g.,][]{harrison:1999,stanek:1999,frail:2000,frail:2001}. These
orphan  afterglows arise  when  the initial  GRB,  and its  associated
afterglow  light, are  directed  away from  the  observer.  Here,  the
deceleration of  the relativistic outflow, the  associated decrease in
special-relativistic  beaming, and the  hydrodynamic spreading  of the
collimated  jet  combine at  the  jet-break  time,  $t_{\rm jet}$,  to
irradiate    a    rapidly    increasing    fraction   of    the    sky
\citep[e.g.,][]{rhoads:1999}.    Observers  illuminated   during  this
transition  will  detect a  steeply  rising  ($\Delta t\approx  t_{\rm
  jet}/10$)  transient that proceeds  to behave  as a  post-jet break,
on-axis  afterglow.    It  will   fade  as  a   power-law  ($F_\nu\propto
t^{-\alpha}$)   with   $\alpha\approx    2.3$,   referenced   to   the
(unconstrained) burst time, resulting  in a decay by $\sim1$\,mag over
a  time scale of  $\Delta t\approx  1.5 t_{\rm  jet}$.  Note  that the
typical observed jet-break times for GRBs are $\sim$1 to 10\,d.


Predicted event rates for PTF are a function of the estimated GRB rate
at low  redshift \citep{guetta:2005,nakar:2006}, the  observed optical
afterglow luminosity distribution \citep{kann:2008}, and the estimated
beaming  (or jet break)  distributions \citep{guetta:2005,nakar:2006}.
There are substantial uncertainties  in each of these functions; thus,
even upper  limits from well-defined orphan afterglow  searches can be
used to constrain interesting properties of the GRB population.

The  discovery  of an  orphan  afterglow  would  serve as  a  dramatic
confirmation  of the ``jet  model'' for  GRBs.  Multiple  events would
begin to  map out the  beaming distribution and, in  addition, provide
inputs to physical models  of relativistic outflows.  Perhaps the most
intriguing possibility relates to  the merging compact object model of
the  short  bursts \citep{fox:2005}.   Detectable  short-burst  orphan
afterglows  will arise  from a  relatively  nearby ($D\lessim20$\,Mpc)
merger event,  potentially within range of  current gravitational wave
detectors.   By providing  a  temporal window  to  searches, and  thus
increasing  their sensitivity,  the discovery  of an  orphan afterglow
could  even in  the  best case  provoke  a first  direct detection  of
gravitational waves \citep{nakar:2006}.

Rapidly evolving ``no-host''  transients identified in previous orphan
afterglow searches \citep{becker:2004,rau:2006,malacrino:2007} are now
attributed  to flare-star  activity \citep{kulkarni:2006}.   The upper
limits on event  rates derived from these surveys  are consistent with
an expectation  for detection of a  few orphan afterglows  per year in
PTF survey work, assuming the  foreground ``fog'' of flare star and CV
activity \citep{rau:2007b} can be successfully penetrated.

%
%


\subsection{Transiting Exoplanets Around Young Stars}
\label{sec:orion}

The  majority of the  $\sim$300 currently  known exoplanets  have been
found  around relatively  old  stars \citep[1--7\,Gyr;][]{saffe:2005}.
In contrast, little  is known about the distribution  and frequency of
planets  orbiting  stars  of  1--100\,Myr  age.   These  objects  have
broadened spectral  lines ($>200-500$\,m s$^{-1}$),  which complicate,
and make more difficult,  the discovery of planets via radial-velocity
variations.   Here, the  detection  of planetary  transits  is a  more
promising method.

Observations of  transiting planets  allow one to  determine important
physical  characteristics  and place  significant  constraints on  the
internal structure of planets \citep{burrows:2000}.  Specifically, the
radius  (obtainable only  for  transiting planets)  provides a  direct
estimate   of   the   age,   composition,  and   surface   temperature
\citep{saumon:1996}.   Studies  show  the  existence  of  ``inflated''
planets with  unexpectedly large  radii or low  densities \citep[e.g.,
WASP-1b,  HAT-P-1b;][]{bakos:2006,cameron:2007}  that  current  models
fail to  explain \citep{pont:2007}.  Next,  assessing the distribution
and  frequency of  planets around  young stars  can also  set detailed
constraints on the time scales of formation and evolution of planetary
systems \citep[e.g.,][]{baraffe:2003}.

\begin{figure}[htb]
\includegraphics[angle=90,width=0.53\textwidth,keepaspectratio=true]{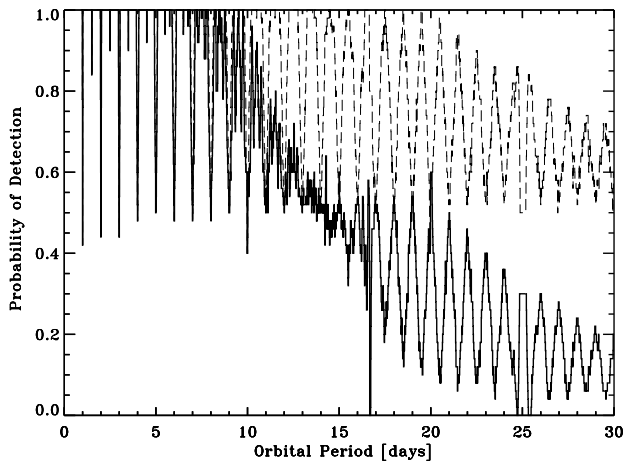}
\caption{Prediction of the window function for the Orion program.  The
  dashed and  solid lines correspond  to the detection  probabilities of
  one  and two transits,  respectively.  The  white (photon)  noise is
  assumed to be  0.01\,mag, and the red (systematic)  noise is assumed
  to be  10\% of the  total photometric noise. Other  input parameters
  are stellar/planetary radius  = 1.0/0.1\,R$_{\odot}$, 40 consecutive
  nights, and a 1\,min observing cadence.}
\label{fig:orion_det}
\end{figure}

During 40  consecutive nights in each  of the first  three years after
commissioning,   PTF   will   provide   high   photometric   precision
measurements  (0.1--2\%)  for  $\sim50,000$  stars  toward  the  Orion
region.   Approximately 5--10\%  of these  stars will  be  young stars
\citep[1--10\,My;][]{carpenter:2001}.   The large number  of dedicated
consecutive nights results in  a $\gtrsim 90$\% sensitivity to orbital
periods  of $P_{orb}\lesssim 10$\,d  (see Figure~\ref{fig:orion_det}).
If  the  fraction  of   the  systems  with  favorable  inclination  is
$\sim10$\% and the planetary distribution and frequency are similar to
those of  main-sequence stars \citep[1 out  of 150;][]{marcy:2004}, we
anticipate 10--15 transiting planets to be found with this experiment.
However,  even a  null  detection  would set  valuable  limits on  the
presence  of  giant planets  with  short  period  orbits. The  transit
search is based upon a box-shaped matched filter algorithm and uses
frequency filtering  and variability-fitting to reduce  the effects of
the  intrinsic variability  of  the stars  \citep[e.g.,][]{aigrain07}.
These techniques  have been very  successful in finding  small transit
signals     in    intrinsically     variable     stars    \citep[e.g.,
CoRoT-7b;][]{leger09}.

The  transit survey  is  most  sensitive to  Jupiter  sized (and  mass)
planets in  few day orbits around stars  of $\lesssim 1$\,M$_{\odot}$.
Follow-up observations will include high spatial resolution imaging to
search  for  background  eclipsing  binaries,  and  medium  resolution
spectroscopy  and radial velocity  monitoring ($\sim  500$\,m s$^{-1}$
precision)  to rule  out stellar  or brown  companions.  Jupiter-sized
planets in short orbital periods produce radial velocity variations on
the  order of  a  few  hundred meters  per  second.  Higher  precision
monitoring ($50  - 200$\,m s$^{-1}$),  coupled with the  known orbital
period  determined from the  photometric observations,  should provide
sufficient  precision  to  confirm  the  presence or  absence  of  the
candidate planets.

The  Orion observations  will  also  provide a  unique  dataset for  a
variety  of  aspects  of  stellar  astrophysics.   The  high-precision
time-series data will enable  the search for, and characterization of,
eclipsing  binary  systems suitable  for  testing  star formation  and
evolution models, characterizing the  activity and rotation periods of
young  stars, and  identifying and  characterizing  previously unknown
young stars in the Orion region.

\subsection{Eclipsing Objects Around Cool Stars}
\label{sec:mdwarfs}

Current radial-velocity surveys  for extrasolar planets preferentially
target bright FGKM stars with masses  within a factor of two of \msun.
Thus  far only $\sim$10  planetary systems  have been  detected around
stars with masses $< 0.5\rm{M_{\odot}}$ and radial-velocity surveys of
M-dwarfs  have  been   limited  to  the  brightest  M0   to  M3  stars
\citep[e.g.,][]{Endl:2006aa,Johnson:2007aa}.  The higher-mass M-dwarfs
that  have  been  probed  have  a low  Jupiter-mass  planet  companion
frequency   \citep[$<\sim$2\%;   eg.][]{Johnson:2007aa},  but   theory
predicts a  large population of  lower-mass planets \citep{Ida:2005aa,
  Kennedy:2008aa}.   The  characteristics  of the  lower-mass  M-dwarf
planet population  are essentially  unknown, mostly because  the stars
are  too  faint  for  current  radial-velocity  planet  searches.   As
M-dwarfs are the most common stars in our galaxy, there is a clear gap
to be addressed in our knowledge of the galactic planetary population.

The PTF R-band DyC experiment offers an opportunity to search M-dwarfs
for stellar eclipses and planetary transits. The survey is expected to
greatly  increase the  number  of known  eclipsing  cool star  systems
(vital for  calibration of the mass-to-luminosity relation  at the low
end),  to  obtain  detailed  statistics  on the  population  of  close
substellar  companions, to  watch  for variability  and activity,  and
ultimately to be capable of detecting low-mass transiting planets.

By  specifically  targeting  faint  M-dwarfs, we  greatly  reduce  the
normally   extreme   precision   requirements   for   planet   transit
surveys.  Going down  to $\rm{m_R}$=20.0,  the faintest  stars  in the
survey will have approximately  SNR=10 per image, sufficient to detect
planets as  small as  Saturn (0.087$\,\rm{R_{\odot}}$) around  M5 dwarfs
($\sim$0.2$\,\rm{R_{\odot}}$)  with  SNR=2  in each of many observations  during
transits. With  few-percent photometric precision  on bright M-dwarfs,
PTF could detect Neptune-radius planets.

In each R-band high-cadence PTF field  we will be able to search a few
thousand M-dwarfs for transitory flux decreases, depending on galactic
latitude. By covering a very  large sample of stars, but at relatively
low photometric precision and  with observation gaps, this survey will
be complementary  to targeted  cool star transit  surveys such  as the
MEarth project \citep{Irwin:2009aa}.

Assuming a representative DyC observing  setup, we expect to search approx.  $\rm{10^5}$ M-dwarfs per year for companions in a variety
of orbits. Scaling against other M-dwarf eclipsing binary surveys, the
survey could  find tens  of new M-dwarf  eclipsing binary  systems per
year. Assuming  that 2\% of  M-dwarfs have giant  planetary companions
with  periods $<$30\,d,  in a  uniform  period distribution,  and
modeling  detection  probability  factors  from the  survey  cadences,
transit  lengths, weather  losses and  the inclination  ranges  of the
possible systems,  PTF also has  the possibility of  detecting several
bona-fide transiting  planet systems per year. Follow  up efforts will
include extensive photometric  and spectroscopic observing programs to
eliminate  false detections,  and ultimately  infrared radial-velocity
measurements  (using,  for  example,  T-EDI  on the  Palomar  Hale  5m
Telescope; Muirhead et al. (2008))   to  confirm  the   presence  of
planets.

\subsection{RR~Lyrae Stars}
\label{sec:rrlyrae}

RR~Lyrae are short-period, yellow  or white giant pulsating variables.
They  are  the most  well-calibrated,  near-field distance  indicators
known.  With a mean  magnitude of $M_R\approx0.3\pm0.2$\,mag for stars
with  $-2.6<$[Fe/H]$<-1.6$\,dex  and  a weak  metallicity  dependence,
RR~Lyrae  have been  used to  measure distances  throughout  the Local
Group.   They have  also been  used to  probe the  kinematics  and the
global     density    distribution     of     the    Galactic     halo
\citep{vivas:2003,ivezic:2004,kinman:2007}.  Having  many thousands of
RR~Lyrae stars over  a significant fraction of the  sky will enable us
to  better constrain not  only the  stellar density  as a  function of
radius, but also the three-dimensional shape of the Galactic halo.

\begin{figure*}[htb]
\begin{center}
\vspace{-0.3in}
\centering
\includegraphics[width=0.95\textwidth,angle=0]{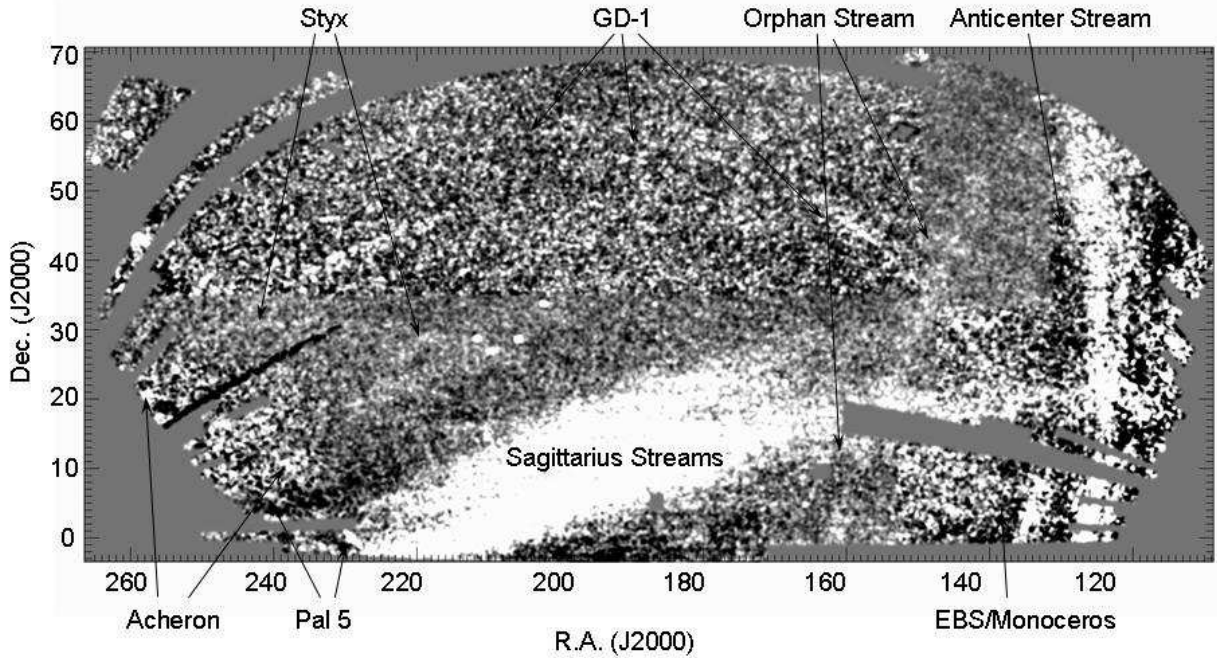}
\end{center}
\vspace{-0.9in}
\caption{A  composite, filtered surface  density map  of stars  in the
  SDSS  Data Release  5. Stars  in DR5  have been  filtered  to select
  stellar  populations  at  different distances  with  color-magnitude
  sequences   similar  to   that  of   the  globular   cluster   M13
  \citep[e.g.,][]{grillmair:2009}.  Lighter shades  indicate  areas of
  enhanced surface  density, and different portions of  the field have
  been filtered for stars at different distances. Varying noise levels
  are  a  consequence  of  the  very different  levels  of  foreground
  contamination using  these different  filters. The distances  of the
  streams range  from 4\,kpc for Acheron,  to 9\,kpc for  GD-1 and the
  Anticenter Stream, to $\sim$50\,kpc for Sagittarius and Styx.}
\label{fig:tidalstreams}
\end{figure*}

Another  strong  motivation  for  extending  the  sample  of  RR~Lyrae
variables is the mapping of substructures within the halo. It is still
not completely clear whether the  Galaxy formed primarily in a single,
homogeneous collapse \citep{eggen:1962}, or the majority of stars were
assimilated  through  the  accretion  of  many  smaller  galaxies  and
galactic fragments \citep{searle:1978}.   The distribution of RR~Lyrae
candidates in the SDSS is decidedly inhomogeneous, and a population of
RR~Lyrae stars  associated with the Sagittarius tidal  stream has been
found  \citep{ivezic:2004}.   Since then,  several  other large  tidal
streams believed  to be  the remnants of  accreted galaxies  have been
detected      out       to      $\sim50$\,kpc      \citep[][;      see
Figure~\ref{fig:tidalstreams}]{yanny:2003,              belokurov:2006,
  grillmair:2006a,  grillmair:2006b,  grillmair:2009}.   A  large-area
survey for RR~Lyrae stars will  allow much more accurate distances and
orbital trajectories  to be  determined for these  streams.  Moreover,
with a  clean sample  of RR Lyrae stars selected through  multi-epoch light
curve analysis, this  would enable us to extend  our reach well beyond
the  $\approx$50\,kpc achieved  to date  using turn-off  stars  in the
SDSS, and  detect additional debris streams  out to $\sim$100\,kpc.
Follow-up  spectroscopy  and  proper-motion measurements  of  RR~Lyrae
stars in  streams will put strong  constraints not only  on the global
Galactic potential but also on its lumpiness \citep{murali:1999}.

Most known RR~Lyrae stars  are fundamental- mode pulsators with periods
on the order of $\lessim 1$\,d. By virtue of its large areal coverage,
the  5DC  will  be  the  most  useful  for  detecting  RR~Lyrae  stars
throughout the  Galactic halo.  While the cadence  is not particularly
adapted  to the detection  of these  stars on  short time  scales, the
$\sim400$  separate images  per field  over the  course of  the survey
(half of  which will  be separated by  $\sim60$\,min) will  enable the
identification  of  RR~Lyrae stars  and  their  mean magnitudes  using
standard time-series filtering techniques.

With a limiting  magnitude per exposure of $R=21.0$,  PTF will be able
to  detect RR~Lyrae  stars at  distances of  10--115\,kpc.   Kinman et
al. (2007)  found 26  RR~Lyrae stars within  8\,kpc in  a 200\,deg$^2$
area centered on the north Galactic  pole.  Using this as a measure of
the local volume density and  adopting a radial density profile in the
Galactic    halo    which     falls    as    $\rho\propto    R^{-2.7}$
\citep{siegel:2002}, an  integration over the  10,000\,deg$^2$ area of
the 5DC yields  an expected total of 28,000  RR~Lyrae stars.  Changing
the   power-law   index  to   $-2.5$   \citep{robin:2000}  or   $-3.3$
\citep{sommer:1990}  yields  estimated totals  of  38,000 and  14,000,
respectively. These numbers are at  least an order of magnitude larger
than any previous RR~Lyrae study;  the final PTF database will clearly
be  a boon  to  investigations of  both  the global  structure of  the
Galactic halo and of the substructures within it.

\subsection{Other Galactic variables}
\label{sec:otherVars}

Beside  the already  mentioned RR~Lyrae  stars many  more  classes and
subclasses of  variables can be  studied with PTF. In  particular, the
combination of PTF variability information and SDSS photometric colors
can  characterize stellar  variability  across the  Hertzsprung-Russel
diagram.  Characterizing  the  detailed  stellar variability  is  also
imperative in the search of genuine transients.

Understanding the  processes that lead to luminosity  changes in stars
is vital to decipher the  stellar evolution.  This 5DC survey with its
wide-area coverage will  enable the discovery of rare  sources such as
R~CrB  stars \citep{iben:1996}, the  non-radial magnetic  pulsating Ap
stars \citep{shibahashi:1987},  and SX~Phe  stars.  The nature  of the
later is  uncertain as they  can be the  result of stellar  mergers or
stars evolving towards the white dwarf stage \citep{cacciari:1990}.

R~CrB  stars are  also  linked to  stellar  mergers, specifically  the
merger of  hydrogen-deficient stars or  of a Neutron star  with Helium
rich  star.   The  birthrate  of   R~CrB  stars  is  estimated  to  be
0.004--0.1\,yr$^{-1}$  in  the  Milky  Way \citep{iben:1996}  and  the
predicted  short  life time  leads  to  estimates  that approx.   1000
sources are  currently existing in  our galaxy. Using their  space and
velocity  distribution,  PTF  may  able  to  resolve  the  fundamental
question  to  which  population (e.g.,  thick  disk)  they belong.

We will construct  light curves for all stars in  the PTF footprint in
the range 14$<$R$<21$\,mag.   Each of these data sets  will consist of
approximately 200-400 data points over  all time scales from about one
hour (and in some cases minutes) to several years.

\subsection{Microlensing}
\label{sec:micro}

Microlensing events  occur when a large stellar-mass  object passes in
front of a background star.  They  take the form of a magnification of
the background star and last  on the order of several weeks, depending
on the  lensing geometry and  lens mass.  Lensing light  curves having
high  temporal  resolution offer  the  opportunity  to detect  planets
around the lensing stars, while  low-mass brown dwarfs and other faint
nearby  objects can be  found by  the lensing  they induce  without an
obvious  lens  star.   Current   microlensing  surveys  such  as  OGLE
\citep[e.g.,][]{pont:2008} and  MOA \citep[e.g.,][]{bennett:2008} find
$\sim$1000 events yr$^{-1}$.

Microlensing  surveys  are  usually  targeted  toward  highly  crowded
stellar  fields  to maximize  the  event  probability.   PTF plans  to
avoid  these fields  to  minimize  stellar contamination  to
extragalactic  transients.   However,  PTF's  much larger  FoV  allows
coverage  of  many more  high  Galactic  latitude  stars than  current
microlensing  projects  can  provide.   On  the other  hand,  at  high
Galactic latitude the number of suitable distant stars to be lensed is
lower.

\cite{han:2008}  has performed  detailed simulations  of  the expected
event  rate for  a  PTF-like  survey.  With  a  limiting magnitude  of
$V=18$\,mag an  all-sky event rate of  $\sim23$\,yr$^{-1}$ is expected
(for events with magnification  $>$1.34).  Approximately half of those
events will occur at $|b|>30$\degs.  Extrapolating the \cite{han:2008}
results to  $R=21$\,mag suggests 200--300  events yr$^{-1}$ throughout
the sky.

With a typical time scale of  about 20\,d, these events can be detected
in  the 5DC with  sufficient time  to organize  follow-up observations
with high temporal resolution for  the later parts of the light curve,
to  search for  planets. Taking  into  account the  PTF footprint,  we
expect the  5DC to find about  5 stellar-stellar events  per year.  In
addition, PTF will be sensitive to microlensing events caused by other
optically faint, massive objects.

\begin{table*}[ht]
\caption{Properties and Rates for Selected Transients and Variables}
\begin{footnotesize}
\begin{center}
\begin{tabular}{lcccccl}
\hline
\hline
Class           &  M$_R$ & $\tau$\tablenotemark{a} & Universal Rate (UR) & PTF Rate & Reference\tablenotemark{b} \\
& [mag] & [days] &  & [yr$^{-1}$] &  $M_R$/$\tau$/UR \\
\hline
Dwarf Novae & 9..4 & 3..20 & $3\times10^{-5}$\,pc$^{-3}$ yr$^{-1}$& 100 &  1/2/3\\
Classical novae & $-5..-10$ & 2..100 & $2\times10^{-10}$\,yr$^{-1}$ L$_{\odot,K}^{-1}$ & 60..150 & 4/5/6 \\ 
Luminous red novae & $-10..-14$ & 20..60 & $1.5\times10^{-13}$\,yr$^{-1}$ L$_{\odot,K}^{-1}$ & 1.5 & 7/7/8\\
Fallback SNe    & $-4..-21$ & $0.5..2$ & $10^{-13}$\,yr$^{-1}$ L$_{\odot,K}^{-1}$  &  1 & 9/9/9\\
Macronovae      & $-13..-15$ & 0.3..3 & $10^{-4..-8}$\,Mpc$^{-3}$ yr$^{-1}$ &  0.1 &  10/10/11\\
SNe~.Ia       & $-15..-17$ & 2..5 & $(4..10)\times10^{-6}$\,Mpc$^{-3}$ yr$^{-1}$ &  0.25..2  &   12/12/12\\
SNe~Ia          & $-17..-19.5$ & 30..70 & \tablenotemark{c}$3\times10^{-5}$\,Mpc$^{-3}$ yr$^{-1}$ & 500 & 13/13/14\\
Tidal disruption flares &$-15..-19$ & 30..350 & $10^{-6}$\,Mpc$^{-3}$ yr$^{-1}$ & 3 & 15/15/15 \\ 
Core-collapse SNe & $-14..-21$ & 20..300 & $5\times10^{-5}$\,Mpc$^{-3}$ yr$^{-1}$  & 200 &  16/*/17\\
Luminous SNe & $-19..-23$ & 50..400 & $10^{-7}$\,Mpc$^{-3}$ yr$^{-1}$ & $>10$ & 18/18/*\\
Orphan afterglows (SGRB)& $-14..-18$ & 5..15 & $3\times10^{-7..-9}$\,Mpc$^{-3}$ yr$^{-1}$ & $<1$ & */*/* \\
Orphan afterglows (LGRB)& $-22..-26$ & 2..15 & $3\times10^{-7..-9}$\,Mpc$^{-3}$ yr$^{-1}$ & $<1$ & */*/*\\
On-axis LGRB afterglows & $..-37$ & 1..15 & $4\times10^{-10}$\,Mpc$^{-3}$ yr$^{-1}$ & 0.5 & */19/20\\
\hline
\end{tabular}
\end{center}
\tablenotetext{a}{Time to decay by 2 mag from peak}
\tablenotetext{b}{References for $M_R$, $\tau$ and universal rate: 1: \cite{rau:2007b}, 2: \cite{sterken:2005}; 3: \cite{schwope:2002}; 
4: \cite{shafter:2009}; 5: \cite{dellaValle:1995}; 6: \cite{ferrarese:2003}; 7: \cite{kulkarni:2007}; 8: \cite{ofek:2008}; 9: Fryer, C.L., priv. com.; 10: \cite{kulkarni:2005}; 11: \cite{nakar:2006}; 12: \cite{bildsten:2007}; 13: \cite{jha:2006}; 14: \cite{dilday:2008}; 15: \cite{strubbe:2009}; 16: \cite{richardson:2002}; 17: \cite{cappellaro:1999}; 18: \cite{scannapieco:2005b}; 19: \cite{fox:2003}; 20: \cite{guetta:2005b}; *: see text}
\tablenotetext{c}{Universal rate at z$<0.12$}
\label{tab:transients_rates}
\end{footnotesize}
\end{table*}

\subsection{Solar System Objects}
\label{sec:neo}

PTF will  detect and measure orbits  for many Solar  System bodies. In
particular,  asteroids  will  form  the dominant  variable  foreground
population.  Depending  on ecliptic latitude, the  predicted number of
known    asteroids   brighter    than   $R=20.5$\,mag    ranges   from
$\sim30$\,deg$^{-1}$    ($\beta=0$\,deg)   to   $\sim0.01$\,deg$^{-1}$
($\beta>50$\,deg).

Near-Earth Objects (NEOs) are a  subgroup of asteroids and comets that
have  been nudged by  the gravitational  attraction of  nearby planets
into orbits  that allow them  to enter the Earth's  neighborhood. They
are of  astronomical interest as nearby samples  of planetesimals left
over from the formation of the Solar System.  The Earth was built from
such  planetesimals, and  the  outer layers  of  Earth, including  the
biosphere, have  been strongly effected  by asteroids and  comets that
impacted the early planet after it had solidified.

NEO discovery  and characterization is also critical  to assessing the
impact risk hazard,  as objects of size 1\,km  could have catastrophic
planet-wide consequences.  Here, PTF  will contribute to NEO discovery
and  orbit  determination by  delivering  astrometric and  photometric
measurements comparable  (or even superior)  to the those of  the best
currently-operating surveys  (Table~\ref{tab:neo}).  In this  way, PTF
will  provide  a  valuable   testing  ground  for  NEO  searches  with
Pan-STARRS and LSST.   The PTF NEO survey will  be performed using the
database of  source extractions obtained  by the main  image-reduction
pipeline. Non-stationary objects will be separated from the stationary
objects detected in the 5DC.

Based  on the estimated  population of  km-size near-Earth asteroids
\citep{bottke:2002},  discovery statistics \citep{moon:2008},  and the
PTF   observing   strategy   (which   avoids  the   ecliptic   plane),
approximately 75 discoveries per year  are within reach of the PTF; in
practice,  the  highly experienced  Catalina  and  LINEAR surveys  are
likely  to report most  of these  discoveries first,  so the  main PTF
contribution will be confirmation and astrometry.

\begin{table}[th]
\caption[]{Comparison of PTF with other NEO surveys}
\label{tab:neo}
 \begin{tabular}{lcccccc}
\hline\hline
Survey & D\tablenotemark{a} & Scale & FoV & Exp. & Mag. Limit & Coverage \\
& [m] & [$''$ pxl$^{-1}$] & [deg$^{2}$] & [s] & [mag] & [deg$^2$/hr] \\
\hline
PTF & 1.2 & 1.0 & 7.78 & 60 & 21.0 & 300 \\
LINEAR\tablenotemark{b} & 1.0 & 2.3 & 2.0 & 5 & 19.2 & 1050 \\
Spacewatch\tablenotemark{c} & 1.8 & 1.0 & 0.3 & 150 & 22.6 & 7.2 \\
Spacewatch & 0.9 & 1.0 & 2.9 & 120 & 21.7 & 70\\
CSS\tablenotemark{d} & 1.5 & 1.0 & 1.2 & 60 & 21.5 & 600\\
CSS & 0.7 & 2.5 & 8.1 & 30 & 19.5 & 420\\
CSS & 0.5 & 1.8 & 4.2 & 60 & 19.0 & 92\\
Pan-STARRS1\tablenotemark{e} & 1.8 & 0.3 & 3.0 & 30 & 24.0 & 360 \\ 
\hline
\tablenotetext{a}{Telescope diameter}
\tablenotetext{b}{Lincoln Laboratories Neart Earth Asteroid Research}
\tablenotetext{c}{http://spacewatch.lpl.arizona.edu/}
\tablenotetext{d}{Catalina Sky Survey; http://www.lpl.arizona.edu/css/}
\end{tabular}
\end{table}  

\subsection{Transients Discovered Outside the Optical Band}
\label{sec:too}

Most   of  the   time,  PTF   will   operate  in   survey  mode   (see
\S~\ref{sec:strategy}).  However, the  7.9\,deg$^2$ field of view
is  also  ideal  to  search  for optical  counterparts  to  transients
discovered  at other  wavelengths (e.g.,  radio, $\gamma$-rays)  or by
other means than electromagnetic radiation (e.g., gravitational waves,
neutrinos, ultra-high-energy cosmic-rays).  The first direct detection
of   gravitational  waves  is   expected  to   come  from   the  Laser
Interferometer Gravitational-Wave  Observatory (LIGO) in  its enhanced
(start  $\sim2009$) and  advanced ($\sim2013$)  stages.  A  variety of
sources, including nearby supernovae and merging compact objects, will
be found in this way.   With typical positional uncertainties of a few
degrees, only wide-field  instruments like PTF will be  able to search
for optical  counterparts, provide arcsecond  localizations, and allow
estimates of distance.

Theory suggests that a few percent of all stellar collapses to neutron
stars  and  black holes  will  result  in  strong gravitational  waves
\citep{dimmelmeier:2007}.   Models include  accretion-induced collapse
leading to bar-unstable neutron stars \citep{liu:2002}, and stars that
collapse   to   black  holes   with   very   rapidly  rotating   cores
\citep{janka:2007}.   These  events  will  likely be  optically  faint
compared to  standard supernovae, but  still detectable in  the nearby
Universe  with PTF.  Once identified,  the light  curves  will provide
insight into  the explosion  physics as well  as into the  fraction of
radioactive  material  advected   into  the  black  hole.   Similarly,
observations  of  the   expected  dim  supernovae  \citep{li:1998}  or
macronovae \citep{kulkarni:2005} associated with LIGO-detected merging
compact  objects will  immediately lead  to  a trove  of results.   In
particular, the ejected mass is  sensitive to the equation of state of
dense matter and  the type of binary (double  neutron star vs. neutron
star/black hole).

The IceCube Neutrino Observatory (Klein et al. 2008) will soon deliver
automated alerts  to follow-up observations of detections  of pairs of
neutrinos. With  typical positional  uncertainties of these  events of
1--2\degs, PTF can help to find the optical counterparts.

PTF will also advance our understanding of the new class of mysterious
radio   transients   which    has   materialized   in   recent   years
\citep[e.g.,][]{levinson:2002,gal-yam:2006,bower:2007}.   These events
were discovered in  archival data; thus, nothing is  known about their
transient appearance at  other wavelengths.  Furthermore, no quiescent
X-ray, optical, near-infrared, and radio counterparts have been found.
The event  rate is  overwhelming and estimated  to be between  100 and
13,000\,deg$^{-2}$\,yr$^{-1}$    with   typical    time    scales   of
$\sim$0.1--1\,d.   Within   the  framework  of   the  dynamic  cadence
experiments, and in combination  with new radio observations, PTF will
attempt to resolve the nature of these events and probe their possible
association  with  the  elusive Galactic population  of  isolated-old
neutron stars \citep{ofek:2009}.

\subsection{Science with Coadded Broad- and Narrow-Band Images}
\label{sec:surveys}

By the  end of 2012, PTF will  have observed each pointing  in the 5DC
footprint  (i.e.,   $\sim10,000$\,deg$^{2}$)  about  60   times.   The
combined $R$-band images will be  deeper than the SDSS $r$-band images
by  about  1--2\,mag.  Although  the  typical  image  quality will  be
2\asec, these  data will  provide an excellent  deep fiducial  sky for
many  astronomical  projects  and  can  be  mined  for  a  variety  of
large-scale science applications.

As  discussed  in section~\ref{sec:strategy},  PTF  will perform  deep
3$\pi$ sr surveys in several narrow-band filters, including H$\alpha$.
The resulting  sky map  will provide an  unprecedented high-resolution
view  on  the structure  of  the  diffuse,  ionized component  of  the
interstellar medium in  the Milky Way. This information  will be vital
for  understanding  the  dynamics  and  evolutionary  history  of  the
interstellar gas. Next, new  structures, such as stellar wind bubbles,
supernova  remnants,  bow-shock   nebulae,  shells  around  old  novae
\citep{shara:2007},  and planetary  nebula shells  with  large angular
extent are anticipated to be found.

Furthermore, we will find low-surface-brightness star-forming galaxies
in the local Universe, and based on the H$\alpha$ emission strength we
will    be    able     to    estimate    the    star-formation    rate
\citep[e.g.,][]{kennicutt:1998} of  the galaxies within  about 40\,Mpc
in our survey footprint (except for galaxies in the zone of avoidance).

\section{Conclusion \& Outlook}
\label{sec:conclusion}

The Palomar  Transient Factory is a  wide-field (7.9\,deg$^2$) imaging
facility at the Palomar 48-inch Oschin Schmidt telescope, dedicated to
mining  the  optical  sky  for  transients and  variables.   Its  main
strengths  are the  exceptionally large  sky  coverage to  a depth  of
$R\approx21.0$\,mag and the almost full-time operation (80\%).  Nearly
real-time discovery of transient candidates will allow rapid follow-up
observations  with a  wide  range of  committed telescopes,  including
optical and near-infrared photometry as well as spectroscopy.

PTF is expected to uncover numerous new members of a variety of source
populations   (see  Table~\ref{tab:transients_rates}),   ranging  from
distant   supernovae   to   nearby  cataclysmic.    Furthermore,   the
exploration of new regions in phase space (e.g., cadence, environment)
often yields  unexpected discoveries.  Thus, PTF is  poised to provide
an  important  contribution  to  filling  the  existing  gaps  in  the
transient phase  space (Figure~\ref{fig:magtau}), paving  the path for
future synoptic surveys, such as LSST.

\acknowledgments 

This paper  is based on  observations obtained with the  Samuel Oschin
Telescope and the 60-inch Telescope at the Palomar Observatory as part
of the  Palomar Transient Factory project,  a scientific collaboration
between the  California Institute of  Technology, Columbia University,
Las  Cumbres Observatory, the  Lawrence Berkeley  National Laboratory,
the  National   Energy  Research  Scientific   Computing  Center,  the
University of Oxford,  and the Weizmann Institute of  Science. SRK and
his   group    were   partially    supported   by   the    NSF   grant
AST-0507734.  J.S.B.  and  his  group were  partially  supported by  a
Hellman  Family Grant, a  Sloan Foundation  Fellowship, NSF/DDDAS-TNRP
grant  CNS-0540352,  and  a  continuing grant  from  DOE/SciDAC.   The
Weizmann Institute  PTF partnership is  supported by an  ISF equipment
grant to AG.  AG's activity is  further supported by a Marie Curie IRG
grant from the EU, and  by the Minerva Foundation, Benoziyo Center for
Astrophysics, a research grant  from Peter and Patricia Gruber Awards,
and the  William Z.  and  Eda Bess Novick  New Scientists Fund  at the
Weizmann  Institute.  EOO  thanks partial  supports from  NASA through
grants    HST-GO-11104.01-A;     NNX08AM04G;    07-GLAST1-0023;    and
HST-AR-11766.01-A.   A.V.F.  and  his group  are grateful  for funding
from NSF  grant AST--0607485, DOE/SciDAC  grant DE-FC02-06ER41453, DOE
grant  DE-FG02-08ER41563,  the TABASGO  Foundation,  Gary and  Cynthia
Bengier, the  Sylvia and Jim  Katzman Foundation, and the  Richard and
Rhoda Goldman Fund.  S.G.D.  and A.A.M.  were supported in part by NSF
grants AST-0407448  and CNS-0540369, and also by  the Ajax Foundation.
The  National Energy  Research Scientific  Computing Center,  which is
supported by the  Office of Science of the  U.S.  Department of Energy
under Contract No.  DE-AC02-05CH11231, has provided resources for this
project by supporting staff  and providing computational resources and
data storage.  L.B.'s research is  supported by the NSF via grants PHY
05-51164  and AST  07-07633. MS  acknowledges support  from  the Royal
Society and the University of Oxford Fell Fund.

\bibliographystyle{apj}




\clearpage



\end{document}